\documentclass[onecolumn,showpacs,preprint,preprintnumbers,amsmath,amssymb]{revtex4}
\usepackage{amssymb}
\usepackage[dvips]{graphicx}
\begin{document}

\title{ Phase separation of electrons strongly coupled with phonons in
cuprates and manganites.}
\author{A. S. Alexandrov}

\affiliation{Department of Physics, Loughborough University,
Loughborough LE11 3TU, United Kingdom\\}

\begin{abstract}

Recent advanced Monte Carlo simulations have not found
superconductivity and phase separation in the Hubbard model with
on-site repulsive electron-electron correlations. We argue that
microscopic phase separations in cuprate superconductors and
colossal magnetoresistance (CMR) manganites originate from a strong
electron-phonon interaction (EPI) combined with unavoidable
disorder. Attractive electron correlations, caused by an almost
unretarded EPI, are  sufficient to overcome the direct inter-site
Coulomb repulsion in these charge-transfer Mott-Hubbard insulators,
so that low energy physics is that of small polarons and small
bipolarons (real-space electron (hole) pairs dressed by phonons).
They form clusters  localised by disorder below the mobility edge,
but propagate as the Bloch states above the mobility edge. I
identify  the Fr\"ohlich finite-range EPI with optical phonons  as
the most essential for pairing and phase separation in
superconducting layered cuprates. The pairing of oxygen holes into
heavy bipolarons in the paramagnetic phase (current-carrier density
collapse (CCDC)) explains also CMR of doped manganites due to
magnetic break-up of bipolarons in the ferromagnetic phase.   Here I
briefly present an explanation of high and low-resistance phase
coexistence near the ferromagnetic transition as a mixture of
polaronic ferromagnetic and bipolaronic paramagnetic domains due to
unavoidable disorder in doped manganites.

\vspace{0.5cm}

Keywords: electron-phonon interaction, phase separation, bipolarons,
cuprates, manganites

\end{abstract}

 \pacs{71.38.-k, 74.40.+k, 72.15.Jf, 74.72.-h, 74.25.Fy}

\maketitle

\section{Introduction}
\begin{figure}
\begin{center}
\includegraphics[angle=-90,width=1.15\textwidth]{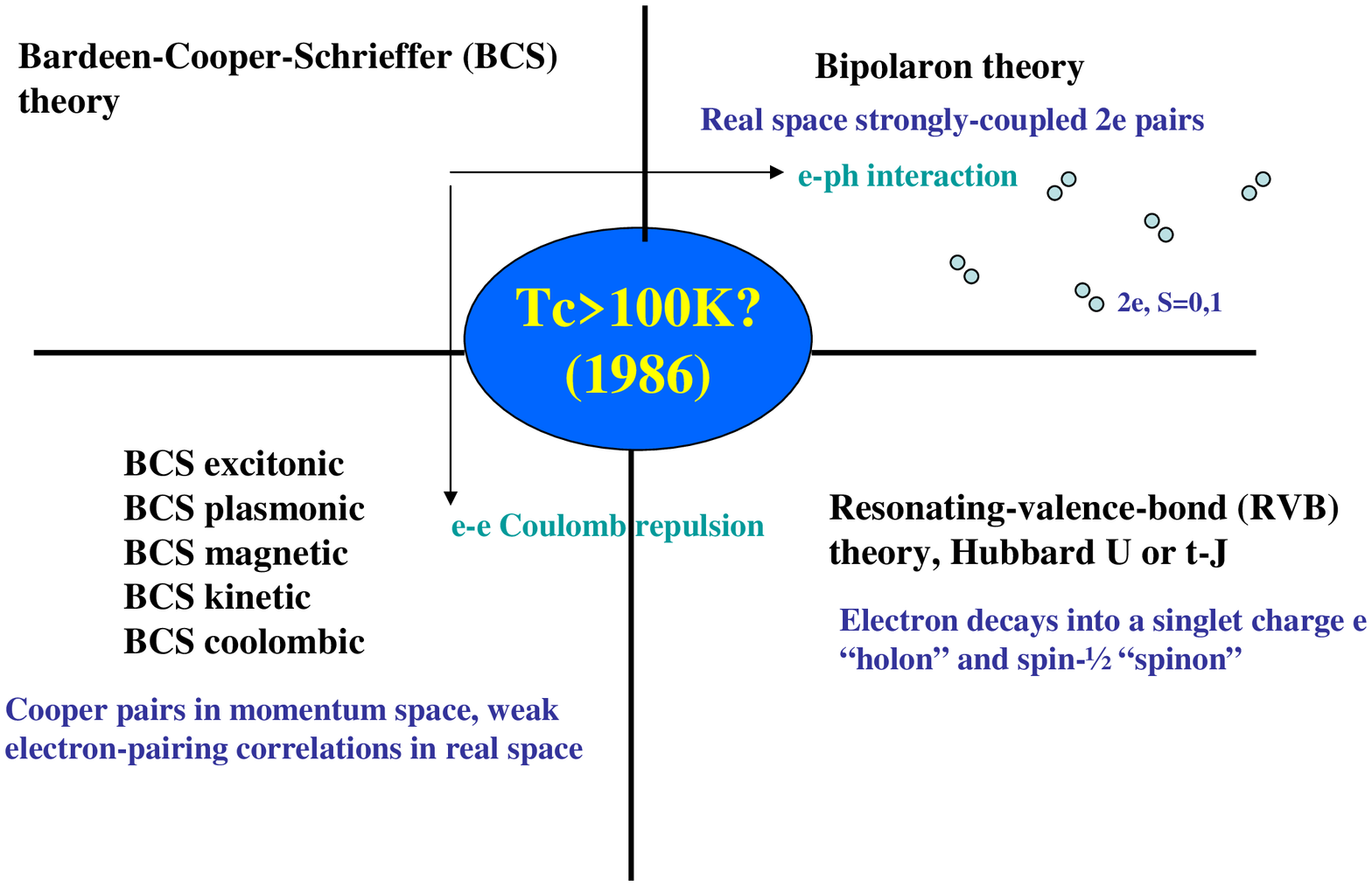}
\vskip -0.5mm \caption{A few theories of high-temperature
superconductivity. The highest $T_c$ is predicted in the BCS to
bipolaron crossover \cite{ale0} for the phonon pairing mechanism
(upper half of the diagram). Lower half of the diagram represents a
number of non-phononic mechanisms of pairing.} \label{dia}
\end{center}
\end{figure}

There are still many theories that attempt to explain the phenomenon
of high-temperature superconductuctivity in cuprates and other
related materials. In general, the pairing mechanism of carriers
 could be not only "phononic" as in the BCS theory \cite{bcs}(left-hand upper corner in Fig.(\ref{dia})) or its
strong-coupling bipolaronic extension \cite{alebook,ale0}
(right-hand upper corner in Fig.(\ref{dia})) , but also "excitonic"
\cite{lit,gin2}, "plasmonic" \cite{fro3,pas},
 "magnetic"  \cite{sch2,mil}, "kinetic" \cite{hir}, or purely "coolombic" due to a mirror-nested Fermi surface \cite{kopaev}.
 The BCS theory like any
mean-field theory is rather universal, so that it describes well the
cooperative quantum phenomenon of superconductivity even with these
non-phononic  mechanisms, if the coupling is weak (left-hand lower
corner in Fig.(\ref{dia})). The main motivation behind these
concepts is that high superconducting critical temperature, $T_{c}$,
could be achieved by replacing phonons in the conventional BCS
theory by higher frequency bosonic modes, such as  plasmons,  spin
waves (pseudomagnons), or even by the direct
 Coulomb repulsion combined with unconventional pairing symmetries.

Actually, following original proposal by P. W. Anderson \cite{and2}
,  many authors \cite{kiv,band} assumed that the electron-electron
interaction in novel superconductors is very strong but repulsive
and  it provides high $T_{c}$ without any phonons (right-hand lower
corner in Fig.(\ref{dia})). A motivation for this concept can be
found in the earlier work by Kohn and Luttinger\cite{kohn}, who
showed that the Cooper pairing of repulsive fermions is possible.
However the same work clearly showed that $T_{c}$ of repulsive
fermions is extremely low, well below the mK scale. Nevertheless,
the BCS and BCS-like theories (including the Kohn-Luttinger
consideration) heavily rely on the Fermi-liquid model of the {\it
normal }state, which fails in many high-temperature superconductors.
If the normal state is not the Fermi-liquid, then there is no direct
reason to reject the assumption. In fact there is little doubt that
strong onsite repulsive correlations (Hubbard $U$) are an essential
feature of the cuprates. Indeed all undoped cuprate compounds are
{\it insulators } with the insulating gap about $2eV$ or so. But if
the repulsive correlations are weak, one would expect a metallic
behaviour of a half-filled $d$-band of copper in cuprates, or, at
most, a much smaller gap caused by lattice and spin distortions
(i.e. due to charge and/or spin density waves \cite{gab,castro}). It
is a strong onsite repulsion of $d$-electrons in cuprates which
results in their parent insulating "Mott"  state. When onsite
correlations are strong and dimensionality is low, there is an
alternative to the usual Fermi-liquid. In Anderson's
resonating-valence-bond (RVB) model \cite {and2} the ground state
supports "topological solitons" (the so-called spinons and holons),
such as occur in one-dimensional Hubbard model. Theoretically holons
could be paired by a superexchange interaction without any
additional glue like phonons or spin-waves \cite{and3}.

To discriminate one theory with respect to another one has to rely
on experimental facts and/or on exact theoretical results. Some
variational Monte Carlo (VMC) simulations with a (projected)
BCS-type trial wave function (see, for example \cite{band} and
references therein), and a number of other analytical and numerical
studies appeared to back up superexchange pairing. However recent
studies by Aimi and Imada~\cite{Imada}, using an advanced
sign-problem-free Gaussian-Basis Monte Carlo (GBMC) algorithm have
shown that these variational methods, as well as other
approximations, overestimated the normal state energy and
 therefore overestimated the condensation energy by at least
an order of magnitude, so that the Hubbard model does not account
for high-temperature superconductivity. The ground state of the
model  is a normal Fermi liquid with no superconductivity and no
stripes. This remarkable result is in line with earlier numerical
studies using the auxiliary-field quantum (AFQMC) \cite{afqmc} and
constrained-path (CPMC) \cite{cpmc} Monte-Carlo methods, none of
which found superconductivity in the Hubbard model.

On the other hand compelling experimental evidence for a strong EPI
has arrived from isotope effects \cite{zhao}, high resolution angle
resolved photoemission spectroscopies (ARPES) \cite{lanzara}, a
number of optical \cite{mic}, neutron-scattering \cite{ega,rez},
tunnelling \cite{davis} spectroscopies of cuprates, and from recent
pump-probe experiments \cite{boz}. I have suggested that
 a strong long-range Fr\"ohlich EPI  is the key to both high-temperature superconductivity and colossal magnetoresistance \cite{alephys}.
Our \cite{ale96,alekor,alekor2,jim2} and some other studies
\cite{Fehske2000,tru,nagaosa} of strongly-coupled polarons and
bipolarons have shown that the long-range discrete Fr\"ohlich EPI in
the presence of the strong Coulomb repulsion does not lead to an
enormous enhancement of the carrier effective mass characteristic of
the Hubbard-Holstein model (HHM), and could provide a higher (room
temperature) superconductivity \cite{alemuller}.

\section{Phase separation in cuprates}

In the  strong-coupling regime, where the BCS electron-phonon
coupling constant is relatively large, $\lambda \gtrsim 1$, pairing
is individual \cite{alebook},   in contrast with the collective
 Cooper pairing \cite{bcs}. Bipolarons survive  even in the normal state above their  Bose-Einstein condensation
 temperature representing a simplest "cluster" of carriers.
While the Fr\"ohlich and Coulomb interactions alone could not lead
to larger clusters like strings or stripes \cite{alekab},
shorter-range interactions as the deformation \cite{kus}, Holstein
\cite{castro2}, Jahn-Teller \cite{muller,kab} or a strong nonlinear
\cite{bus} EPIs could favor bound states of more than two carriers.

Formation of polaronic clusters can be analytically studied in the
strong-coupling regime in the framework of a generic
"Fr\"ohlich-Coulomb" model (FCM) \cite{ale96,alebook,alekor2}. The
model Hamiltonian explicitly includes a long-range electron-phonon
and the Coulomb interactions as well as the kinetic and deformation
energies.  The implicitly present large Hubbard $U$ term prohibits
double occupancy and removes the need to distinguish fermionic spins
since the exchange interaction is negligible compared with the
direct Coulomb and the electron-phonon interactions in complex
oxides.

 Introducing  fermionic, $c_{\bf n}$, and
phononic, $d_{{\bf m}\alpha }$, operators the FCM Hamiltonian   is
written as
\begin{eqnarray}
H = & - & \sum_{\bf n \neq n'} \left[ T({\bf n-n'}) c_{\bf
n}^{\dagger } c_{\bf n'} - {1\over{2}} V_{c}({\bf n-n'}) c_{\bf
n}^{\dagger}
c_{\bf n}c_{\bf n'}^{\dagger } c_{\bf n'} \right]  \nonumber \\
& - &  \sum_{\alpha,\bf n m} \omega_{\alpha} g_{\alpha}({\bf m-n})
({\bf e}_{\alpha } \cdot {\bf u}_{\bf m-n}) c_{\bf n}^{\dagger }
c_{\bf n}
(d_{{\bf m}\alpha}^{\dagger}+d_{{\bf m}\alpha }) \nonumber \\
& + &
 \sum_{{\bf m}\alpha} \omega_{\alpha}\left( d_{{\bf m}\alpha
}^{\dagger} d_{{\bf m}\alpha }+1/2 \right),
\end{eqnarray}
where $T({\bf n})$ is the hopping integral in a rigid lattice, ${\bf
e}_{ \alpha}$ is the polarization vector of the $\alpha$th vibration
coordinate, ${\bf u}_{\bf m-n} \equiv ({\bf m-n})/|{\bf m-n}|$ is
the unit vector in the direction from electron ${\bf n}$ to  ion
${\bf m}$, $g_{\alpha}({\bf m-n)}$ is the dimensionless EPI
function, and $V_{c}({\bf n-n'})$ is the inter-site Coulomb
repulsion. $g_{\alpha}({\bf m-n)}$ is proportional to the {\em
force} $f_{\mathbf{m}}(\mathbf{n})$ acting between the electron on
site ${\bf n}$ and the ion on ${\bf m}$. For simplicity we assume
that all phonon modes are non-dispersive with  frequencies
$\omega_{\alpha}$ and include spin in the definition of ${\bf n}$
(here $\hbar =1$).

\begin{figure}[tbp]
\begin{center}
\includegraphics[angle=-0,width=0.50\textwidth]{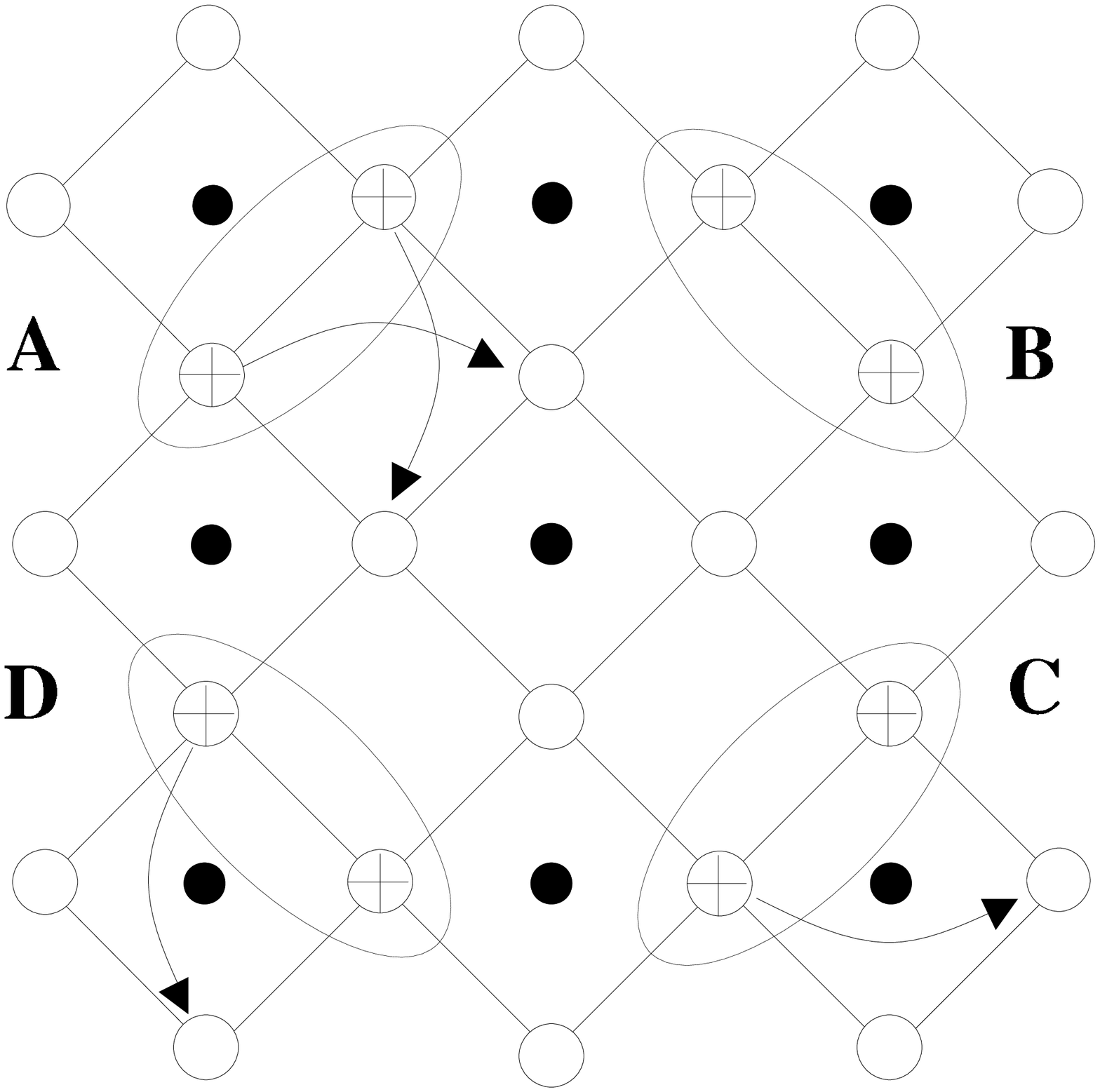} \vskip -0.5mm
\end{center}
\caption{Four degenerate in-plane bipolaron configurations A, B, C,
and D . Some single-polaron hoppings are indicated by arrows.
(Reproduced from \cite{alekor2}, (c) IoP, 2002.) }\label{four}
\end{figure}

Bipolarons on a two-dimensional lattice of ideal octahedra (that can
be regarded as a simplified model of the copper-oxygen perovskite
layer) have been studied in Ref. \cite{alekor2}. Due to poor
screening, the hole-ion interaction was taken as purely coulombic,
\[
g_{\alpha }({\bf m-n})=\frac{\kappa _{\alpha }}{|{\bf m-n}|^{2}},
\]
where $\alpha =x,y,z$ with $\kappa _{x}=\kappa _{y}=\kappa
_{z}/\sqrt{2}$ accounting for the experimental fact that c-axis
($z$-polarized) phonons couple to in-plane holes stronger than
others. The direct hole-hole repulsion is
\[
V_{c}({\bf n-n^{\prime }})=\frac{V_{c}}{\sqrt{2}|{\bf n-n^{\prime
}}|},
\]
so that the repulsion between two holes in the nearest neighbour
(NN) configuration is $V_{c}$. The nearest neighbour hopping
$T_{NN}$, the next-nearest neighbour (NNN) hopping across copper
$T_{NNN}$ and the hopping between the pyramids $T_{NNN}^{\prime }$
have been included, Fig.(\ref{four}).

The polaron level shift in this model is given by the lattice sum,
\begin{equation}
E_{p}=2\kappa _{x}^{2}\omega _{0}\sum_{{\bf m}}\left( \frac{1}{|{\bf m-n}%
|^{4}}+\frac{h^{2}}{|{\bf m-n}|^{6}}\right) =31.15\kappa
_{x}^{2}\omega _{0}, \label{televen}
\end{equation}
where the factor $2$ accounts for two layers of apical sites, and
 the in-plane lattice constant is $a=1$ and
$\omega_{\alpha}=\omega_0$. For reference,
the Cartesian coordinates are ${\bf n}=(n_{x}+1/2,n_{y}+1/2,0)$, ${\bf m}%
=(m_{x},m_{y},h)$, and $n_{x},n_{y},m_{x},m_{y}$ are integers. The
polaron-polaron attraction is
\begin{equation}
V_{ph}({\bf n-n^{\prime }})=4\omega_{0} \kappa _{x}^{2}\sum_{{\bf m}}\frac{%
h^{2}+({\bf m-n^{\prime }})\cdot ({\bf m-n})}{|{\bf m-n^{\prime }}|^{3}|{\bf %
m-n}|^{3}}.  \label{ttwelve}
\end{equation}
Performing the lattice summations for the NN, NNN, and NNN$^\prime$
configurations one finds $V_{ph}=1.23\,E_{p},$ $0.80\,E_{p}$, and
$0.82\,E_{p}$,
respectively. As a result, we obtain a net inter-polaron interaction as $%
v_{NN}=V_{c}-1.23\,E_{p}$, $v_{NNN}=\frac{V_{c}}{\sqrt{2}}-0.80\,E_{p}$, $%
v_{NNN}^{\prime }=\frac{V_{c}}{\sqrt{2}}-0.82\,E_{p}$, and the mass
renormalization exponents (see below) as $g_{NN}^{2}=0.38(E_{p}/\omega)$, $%
g_{NNN}^{2}=0.60(E_{p}/\omega)$ and $(g^{\prime
}{}_{NNN})^{2}=0.59(E_{p}/\omega)$.
 At
$V_{c}>1.23\,E_{p}$, no bipolarons are formed and the system is a
polaronic Fermi liquid. Polarons tunnel in the {\em square} lattice
with the renormalised hopping integrals $t=T_{NN}\exp
(-0.38E_{p}/\omega)$ and $t^{\prime }=T_{NNN}\exp
(-0.60E_{p}/\omega)$ for NN and NNN hoppings, respectively. The
polaron mass is $m^{\ast }\propto 1/(t+2t^{\prime })$.

If $V_{c}<1.23\,E_{p}$, then intersite NN bipolarons form. The
intersite bipolarons
tunnel in the plane via four resonating (degenerate) configurations $A$, $B$%
, $C$, and $D$, as shown in Fig.(\ref{four}). In the first order of
the renormalised hopping integral, one should retain only these
lowest energy configurations and discard all the processes that
involve configurations with higher energies. These inter-site
bipolarons already move in the {\em first} order of the single
polaron hopping. This remarkable property is entirely due to the
strong on-site repulsion and long-range electron-phonon interactions
that leads to a non-trivial connectivity of the lattice. This fact
combines with a weak renormalization of $t^{\prime }$ yielding a
{\em superlight} bipolaron with the mass $m^{\ast \ast }\propto \exp
(0.60\,E_{p}/\omega )$. We recall that in the Holstein model
$m^{\ast \ast }\propto \exp (2E_{p}/\omega )$ \cite{alebook}. Thus
the mass of the small Fr\"{o}hlich bipolaron in the perovskite layer
scales approximately as a {\em cubic root} of that of the Holstein
bipolaron.

At even stronger EPI, $V_{c}<1.16E_{p}$, NNN bipolarons become
stable. More importantly, holes can now form 3- and 4-particle
clusters. The dominance of the potential energy over kinetic in the
transformed Hamiltonian enables us to readily investigate these
many-polaron cases. Three holes placed within one oxygen square have
four degenerate states with the energy
$2(V_{c}-1.23E_{p})+V_{c}/\sqrt{2}-0.80E_{p}$. The first-order
polaron hopping processes mix the states resulting in a ground state
linear
combination with the energy $E_{3}=2.71V_{c}-3.26E_{p}-\sqrt{%
4t^{2}+t^{\prime }{}^{2}}$. It is essential that between the squares
such triads could move only in higher orders of polaron hopping. In
the first order, they are immobile. A cluster of four holes has only
one state within
a square of oxygen atoms. Its energy is $E_{4}=4(V_{c}-1.23E_{p})+2(V_{c}/%
\sqrt{2}-0.80E_{p})=5.41V_{c}-6.52E_{p}$. This cluster, as well as
all bigger ones, are also immobile in the first order of polaron
hopping. Hence a strong EPI  combined with the Coulomb repulsion
could cause clustering of polarons into finite-size  mesoscopic
textures. Importantly QMC
 studies of mesoscopic textures \cite{kab} including lattice deformations
and the Coulomb repulsion show that pairs (i.e. bipolarons) dominate
over phase separation since they effectively repel each other
\cite{alebook}. I would like to stress that at distances much larger
than the lattice constant the polaron-polaron interaction is always
repulsive, and the formation of infinite clusters, stripes or
strings is  prohibited \cite{alekab}.

\section{Phase separation in ferromagnetic manganites}

The conventional double-exchange (DEX) model of the ferromagnetism
and colossal magnetoresistance (CMR), proposed half a century ago
and generalized more recently to include the electron-phonon
interaction, is in conflict with a number of contemporary
experiments \cite{alebra,alebrakab}. Among those experiments are
site-selective spectroscopes, which show that oxygen p-holes are
current carriers, rather than d-electrons in ferromagnetic
manganites \cite{spectr}. Also, some ferromagnetic manganites
manifest an insulating-like optical conductivity at all
temperatures, contradicting the DEX notion that their ferromagnetic
phase is metallic \cite{calvani}.  CMR is observed in manganese
pyrochlores \cite{ramirez}  where DEX is non-existent.

On the other hand, the pairing of oxygen holes into heavy bipolarons
in the paramagnetic phase (current carrier-density collapse (CCDC))
and their magnetic break-up in the ferromagnetic phase is compatible
with the above and many other observations explaining  CMR, isotope
effects, and pseudogaps observed in doped manganites
\cite{alebra2,wang,zhao,alezhao,dessau}. Different from other models
CCDC predicted the first-order phase transition, now firmly
established in single CMR crystals \cite{phillips}. CCDC was
directly observed in the Hall effect at the ferromagnetic transition
\cite{hall}.

\begin{figure}
\begin{center}
\includegraphics[angle=-90,width=1.20\textwidth]{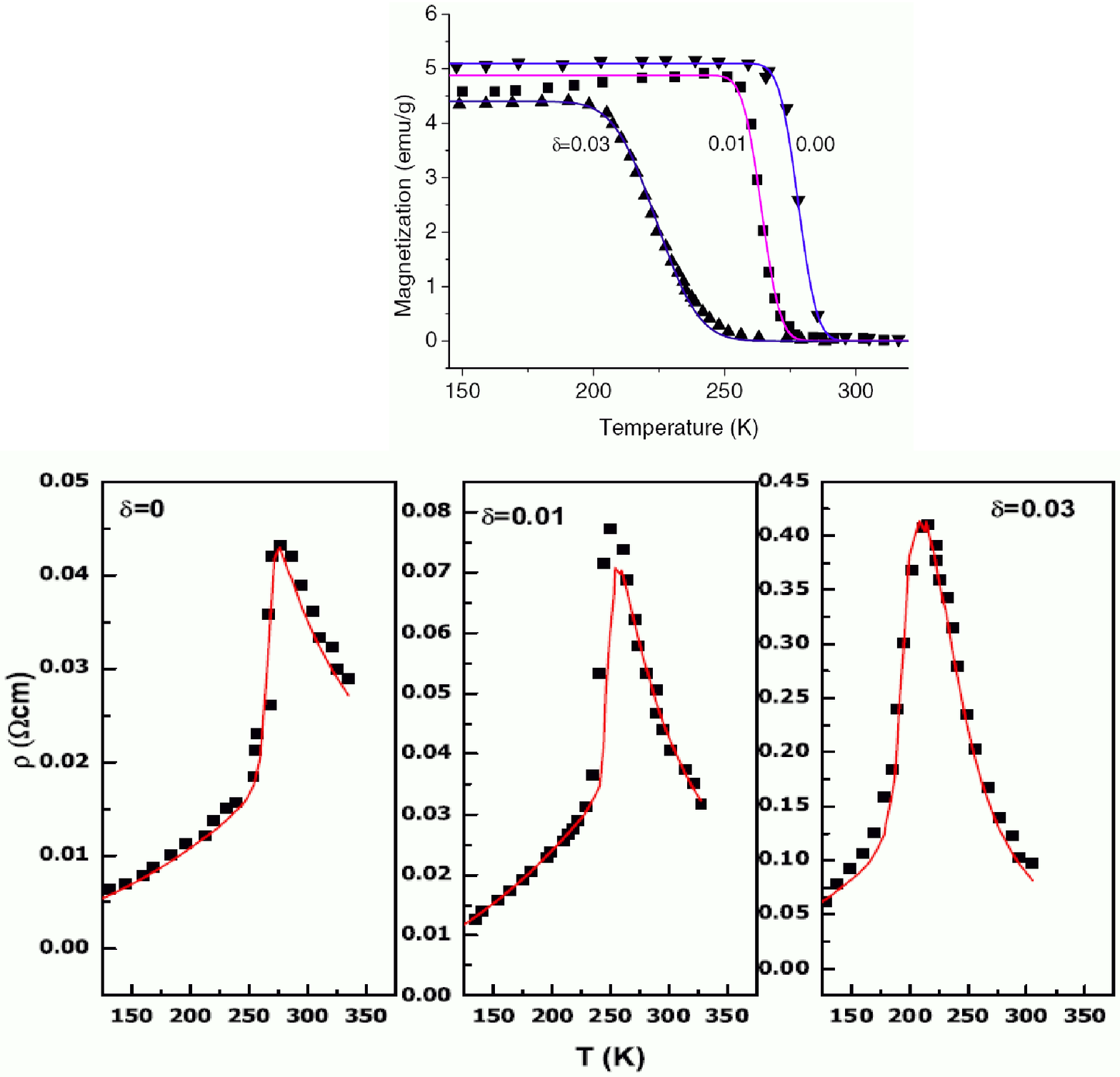}
\vskip -0.5mm \caption{CCDC model ( lines) describes the
experimental magnetisation (upper panel) and
 resistivity (lower panels)  near the ferromagnetic transition
in La$_{0.7}$Ca$_{0.3}$Mn$_{2-\delta}$Ti$_{\delta}$O$_{3}$ (symbols
\cite{china}), if the phase coexistence caused by disorder is taken
into account. No fitting parameters are used in Eq.(\ref{rho}) but
the experimental resistivity well below and well above the
transition and the experimental magnetization. (Reproduced from
\cite{alebrakab}, (c) American Physical Society, 2006.) }
\label{ccdc}
\end{center}
\end{figure}

More recently we have proposed  an explanation of high and
low-resistive phase mixing near the ferromagnetic transition
observed in tunnelling \cite{tun} and  some other experiments as the
mixture of polaronic ferromagnetic domains and bipolaronic
paramagnetic domains due to unavoidable disorder in doped manganites
\cite{alebrakab}. Using the fact that the  phase transition in a
homogeneous system is of the first order in a wide range of
parameters \cite{alebra,alebra2} one can average the magnetisation,
$\sigma(T)$, with the Gaussian distribution of random transition
temperatures $T_{Ci}$s caused by disorder around the experimental
Curie temperature
 $T_C$ to obtain
\begin{equation}
\sigma(T)={1\over{2}} \mathrm{erfc}
\left({{T-T_C}\over{\Gamma}}\right), \label{sigma}
\end{equation}
where $\Gamma$ is the distribution width  and $\mathrm{erfc}(z)=
(2/\pi^{1/2}) \int _z^{\infty} dy \exp(-y^2)$.
 The CCDC with disorder, Eq.~(\ref{sigma})
fits nicely the experimental magnetizations  near the transition
with physically reasonable $\Gamma $ of the order of 10K, depending
on doping, Fig.(\ref{ccdc}). Hence, we believe that the random
distribution of transition temperatures with the width $\Gamma $
across the sample caused by the randomness of the bipolaron binding
energy is responsible for the phase coexistence near the transition
as seen in the tunnelling experiments \cite{tun}.

Resistivity of inhomogeneous two-phase systems has to be calculated
numerically. Nevertheless, the comprehensive numerical simulations
are consistent with a simple analytical expression for the
resistivity of the binary mixture,
\begin{equation}
\rho= \rho_1^{1-\nu}\rho_2^{\nu}, \label{rho}
\end{equation}
which is valid in a wide range of the ratios $\rho _{1}/\rho _{2}$
\cite {kab2}. Here
 $\rho_{1,2}$ is the resistivity of each phase, respectively,
and $\nu$ is the volume fraction of the second phase.

In the framework of CCDC, the resistivity of the paramagnetic phase
is $\rho _{1}(T)=f(T)\exp (\Delta /2k_BT)$ and the resistivity of
the ferromagnetic phase is $\rho _{2}(T)=\phi (T)$, where $f(T)$ and
$\phi (T)$ are polynomial functions of temperature depending on the
scattering mechanisms, and $\Delta$ is the bipolaron binding energy.
Well below the transition $\phi (T)$ can be parameterized  as $\phi
(T)=\rho _{0}+aT^{2}$, and $f(T)=bT$ well above the transition,
where the temperature independent parameters $\rho _{0}$, $a$,
$\Delta/2$ and $b$ are taken directly from the experiment. The
volume fraction $\nu $ of the ferromagnetic phase is simply the
relative magnetization in our model, $\nu =\sigma (T)$, also
available from the experiment. As a result, Eq.~(\ref{rho}) provides
the quantitative description of $\rho (T)$ in the transition region
without any fitting  parameters by using the experimental
resistivity far away from the transition and the experimental
magnetization,  as shown in Fig.(\ref{ccdc}). Studies of the
low-field magnetoresistance of Sm$_{1-x}$Sr$_x$MnO$_3$ (x=0.45)
which was sintered at different elevated temperatures followed by
fast cooling also found  very good qualitative agreement with CCDC
\cite{jung}.

Finally, our concept of polaronic metal in ferromagnetic manganites
\cite{alebra,alezhao} has been substantiated by the angle-resolved
photoemission spectroscopy data for the bilayer manganite
La$_1.2$Sr$_1.8$Mn$_2$O$_7$, where  a polaron metallic state below
$T_C$ has been clearly observed \cite{shenCMR}.

\section{Conclusions}

\begin{figure}
\begin{center}
\includegraphics[angle=-90,width=1.00\textwidth]{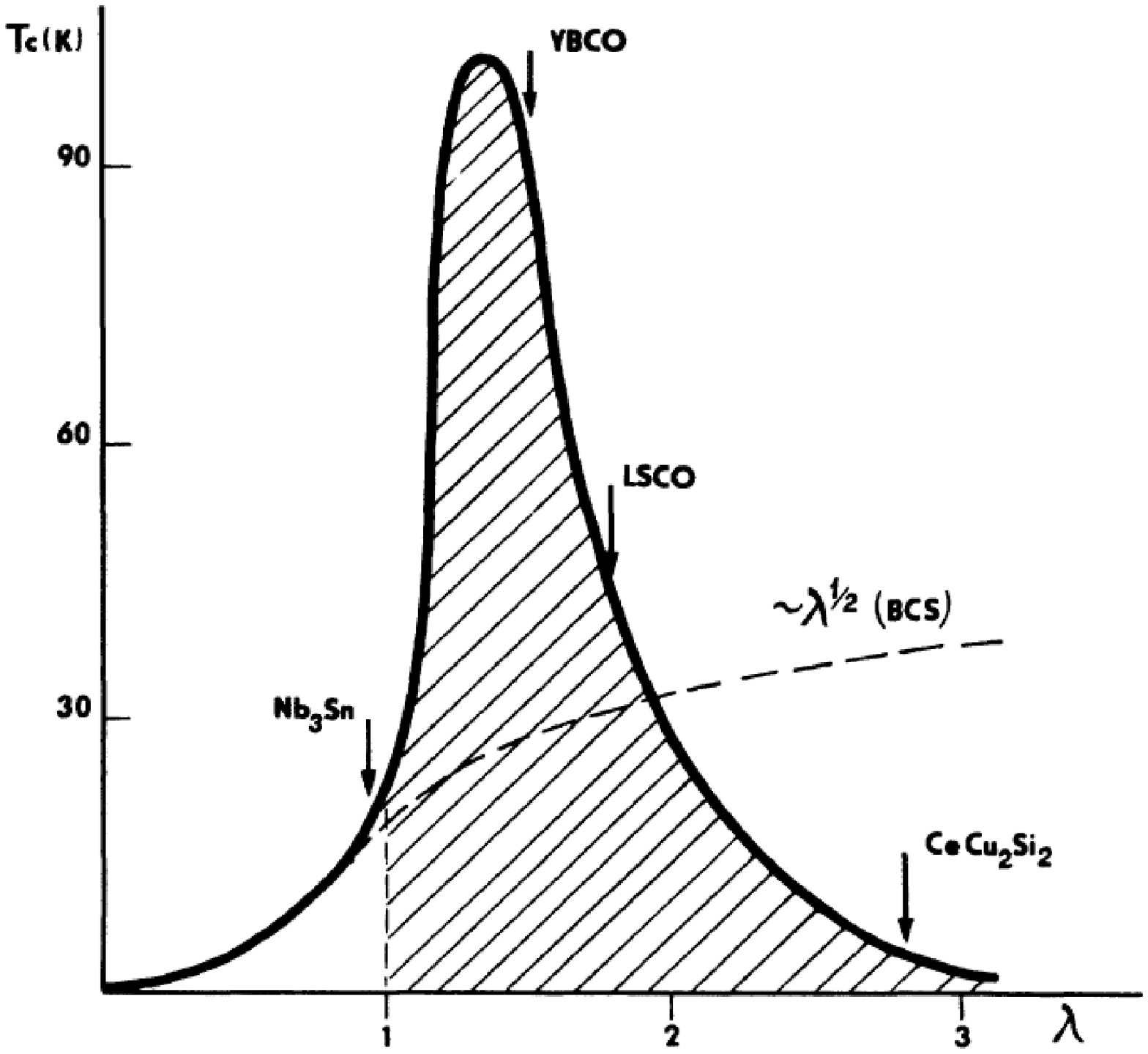}
\vskip -0.5mm \caption{Dependence of the superconducting critical
temperature on  EPI coupling constant. The shading shows the
(bi)polaronic domain. The dotted line corresponds to the
BCS-Eliashberg theory. (Reproduced from \cite{ale1}, (c) American
Physical Society, 1988.)} \label{Tc}
\end{center}
\end{figure}

For although high-temperature superconductivity has not yet been
targeted as `{\it the shame and despair of theoretical physics}', -
a label attributed to conventional superconductivity during the
first half-century after its discovery - the parlous state of
current theoretical constructions has led to a current consensus
that there is no consensus on the theory of high-$T_{c}$
superconductivity. Nevertheless impressive amount of experimental
data (for example \cite{zhao,lanzara,mic,ega,rez,davis,boz}) and
accurate numerical simulations \cite{Imada} have ruled out the
simple Hubbard model as an explanation of high-temperature
superconductivity (lower right-hand corner in Fig.(\ref{dia})). Our
view, which I have briefly presented here in connection with the
phase separation, is that the extension of the BCS theory towards
the strong interaction between electrons and ion vibrations (upper
right corner in Fig.(\ref{dia})) describes the phenomenon naturally.
The high temperature superconductivity exists in the crossover
region of the EPI strength from the BCS-like polaronic  to
bipolaronic superconductivity as was predicted by us \cite{ale0}
before the discovery \cite{muller0}, proposed as an explanation of
high $T_c$ in cuprates \cite{alej,ale1}, Fig.(\ref{Tc}), and
explored in greater detail after the discovery
\cite{rice,emi0,alemot,workshop,dev,alebook}.

Bipolarons also explain CMR \cite{alebra} and, combined with
disorder, the phase separation in manganites \cite{alebrakab}. The
observation of the pseudogap and nodal quasiparticles in colossal
magnetoresistive manganites \cite{shen2}  which have been considered
as a characteristic feature of the copper oxide, further
substantiates  our analogy between high-temperature superconducting
and CMR oxides.

  This work was supported by  EPSRC  (UK) (grant
EP/D035589/1)

\end{document}